\documentclass[reprint, amsmath,
superscriptaddress,
nobibnotes,
nofootinbib,showpacs,
amssymb,aps,prb]{revtex4-1}

\usepackage[dvips]{graphicx}
\usepackage{dcolumn}
\usepackage{bm}
\usepackage{hyperref}
\usepackage{multirow}
\usepackage{array}
\usepackage{color,soul}
\usepackage[textsize=tiny]{todonotes}
\begin{document}

\title{Stability of quasi-two dimensional zigzag carbon and its reaction pathway to graphene}
\author{Andrey Tokarev}
\email[Corresponding Author: ]{tokarev@ualberta.ca}
\affiliation{Department of Mechanical Engineering, University of Alberta, Edmonton, AB, T6G 2G8, Canada}

\author{Bhalchandra S. Pujari}
\affiliation{Present address:Department of Physics and Astronomy, University of Nebraska-Lincoln, Lincoln, NE, 68588-0299. USA}

\begin{abstract}
Using the density functional theory we investigate a quasi-two dimensional
carbon allotrope, ZzC, formed by square carbon lattice buckled in zigzag manner.
By analyzing the Kohn-Sham energy and phonon dispersion obtained by lattice
dynamical calculations we show that ZzC is stable with binding energy of 7.46 eV
per atom. To examine the possible route of formation we find out reaction
pathway from ZzC to graphene using  nudge elastic band method, generalized for
solid state calculations. The reaction pathway shows the formation of carbyne as the intermediate state. Such a pathway is seen to exhibit two transitions states with reaction barriers of 0.21 eV/atom from ZzC to carbyne, and of 1.19 eV/atom from graphene to carbyne. Although ZzC is stable, upon hydrogenation it dissociates and prefers the carbyne-like structure by forming
chains of polyacetylene.   
\end{abstract}

\maketitle

\section{Introduction}
Variety of $sp$ hybridizations of carbon allows it to have diversity of
structures in all three dimensions \cite{CHEM:CHEM200903128}. After discovery of graphene, 2D family of
carbon allotropes became subject of special interest. For example, hydrogenated
graphene, called as {\it graphane} is under investigation by both theoretical
and experimental groups\cite{elias01302009,balog,sofo,PhysRevB.77.035427,
prachi,pujari:graphane, barnard,PhysRevB.81.205417}. Other
materials like graphone and graphYn have also been predicted theoretically. Zhou
{\it et al}\cite{Zhou2009} showed that semihydrogenated graphone can act as a
ferromagnetic material.  On the other hand graphyne has been claimed to be
better than graphene\cite{Baughman,Enyashin,Malko}. An exhaustive theoretical study was carried out recently by Wen {\it et
al}\cite{CHEM:CHEM200903128} who investigated various allotropes of Group 14
elements using Density functional theory (DFT). They studied a variety of 1D, 2D
and 3D structures of group IV elements; particularly their geometries and stabilities
by means of formation energies. Their extensive study shed the light on various
structures and their relative stability.  They also found that in 2D a wavy
square sheet of carbon is considerably more stable than other crystals except
graphene. It thus becomes very intriguing to investigate such wavy material in details,
which we call  Zigzag carbon (ZzC).

In this work we investigate the crystal structure, stability and
electronic structure of ZzC.  The relative stability is investigated by using binding energy (BE) and phonon dispersion spectra as BE cannot be considered conclusive. Moreover, we aslo investigate the reaction pathway from ZzC to graphene. Such a transition
involves both atomic and cell degrees of freedom. We address this transition
using Linear Synchronized Transit (LST), Quadratic Synchronized Transit (QST) \cite{QST} and Generalized Solid-State Nudge Elastic Band (G-SSNEB) \cite{Sheppard2012a} methods. Although LST is easy to implement, QST, and specially G-SSNEB are more accurate in predicting
transition state. We observed that QST reaction barrier from ZzC to carbyne is in 5\% in agreement with G-SSNEB one.

\section{Computational Details}
We perform constraint free full unit cell optimization using Generalized
Gradient Approximation (GGA) with exchange correlation functional of Perdew,
Burke and Ernzerhof \cite{PBE} as implemented in Quantum ESPRESSO \cite{QE-2009}
The primitive cell contains two carbon atoms.  The unit cell optimization was
terminated upon reaching the pressure cut-off of 0.1 Kbar.  Kinetic energy
cutoff on plane waves was set to as high as 1360 eV.  The energy criterion for
electronic self consistency was set to $10^{-8}$ eV while that for structural
optimization was $10^{-7}$ eV.  For the optimization the $k$-mesh was sampled
using $16\times16\times1$ points using Monkhorst-Pack scheme, while the band
structure was plotted on lines joining four symmetry points of Wigner-Seitz cell,
and the individual line segments were sampled using fifty $k$-points. A vacuum
space of 25 {\AA} was kept normal to the  plane to avoid any interplanar
interactions. 

Phonon dispersion is obtained by lattice dynamical calculations performed using
self-consistent density functional perturbation theory within linear response
approach.  The energy threshold value for convergence was 1.4 $\times$
10$^{-15}$ eV.  Force constant matrices are obtained on a 3 $\times$ 3 $\times$
1 $k$-point mesh of irreducible part of Brillouin zone, and are interpolated at
arbitrary wave vectors. The dispersion spectrum is then plotted along the lines
joining symmetry points sampled using total of 150 $q$ points.

For finding solid-state transition pathway between graphene and ZzC we used
three methods: LST, QST and G-SSNEB. In LST all degrees of freedom (DOF), namely
cell parameters and atomic positions, are varied linearly from reactants to
products. The maximum on the resulting pathways corresponds to a transition
state (TS). Better results of TS are obtained by QST, in which one should first
optimize the LST maximum orthogonally to its pathway in combined phase space of
coordinates and cell vectors. (In this work we finally changed orthogonal
optimization to unrestricted optimization because the resulting structure appeared to be another carbon allotrope with energy below ZzC.) Then a quadratic transit
pathway is built from reactants to products through this TS.  For detailed
discussion on QST we refer reader to reference \onlinecite{QST}. The G-SSNEB
method, which we implemented in QE, was developed by Henkelmann's group \cite{Sheppard2012a}. The only
difference with original method is the cell gradient  (Eq.(\ref{eq:cellgrad})), which
we used in the form given by Caspersen and Carter. \cite{Carter} Choice of gradient is based on ease of numerical implementation and
although the nature of gradients are different in both the cases; we believe that they
should produce identical results in a simple case that we are studying.

The gradient in the
combined phase space is the generalized forces, which are
\begin{equation}
	-\frac{\partial E}{\partial \textbf{x}} = \textbf{F $\cdot$} \textbf{h}^T
	\label{eq:corgrad}
\end{equation}
\begin{equation}
	-\frac{\partial E}{\partial \textbf{h}} 
	=  - \Omega \left(\textbf{ ($\sigma$} + P\textbf{I})  \textbf {$\cdot$}   \textbf{h}^{-1}\right)^T  
	\label{eq:cellgrad}
\end{equation}

where \textbf{x}, \textbf{F}, \textbf{h}, \textbf{$\sigma$} are matrices of
atomic positions in crystal coordinates, forces, cell vectors and internal Cauchy stress tensor accordingly.
$P$ is the external pressure, equaled zero in our calculations.

\section{Results and Discussion}
We shall begin the discussion by the presenting the lattice structure and the
stability, and proceed to electronic structure and the reaction pathway from
ZzC to graphene. We shall also talk about the effect of hydrogenation. 
\subsection{Crystal: structure and stability}
\begin{figure*}
	\begin{center}
		\includegraphics[width=.58\textwidth]{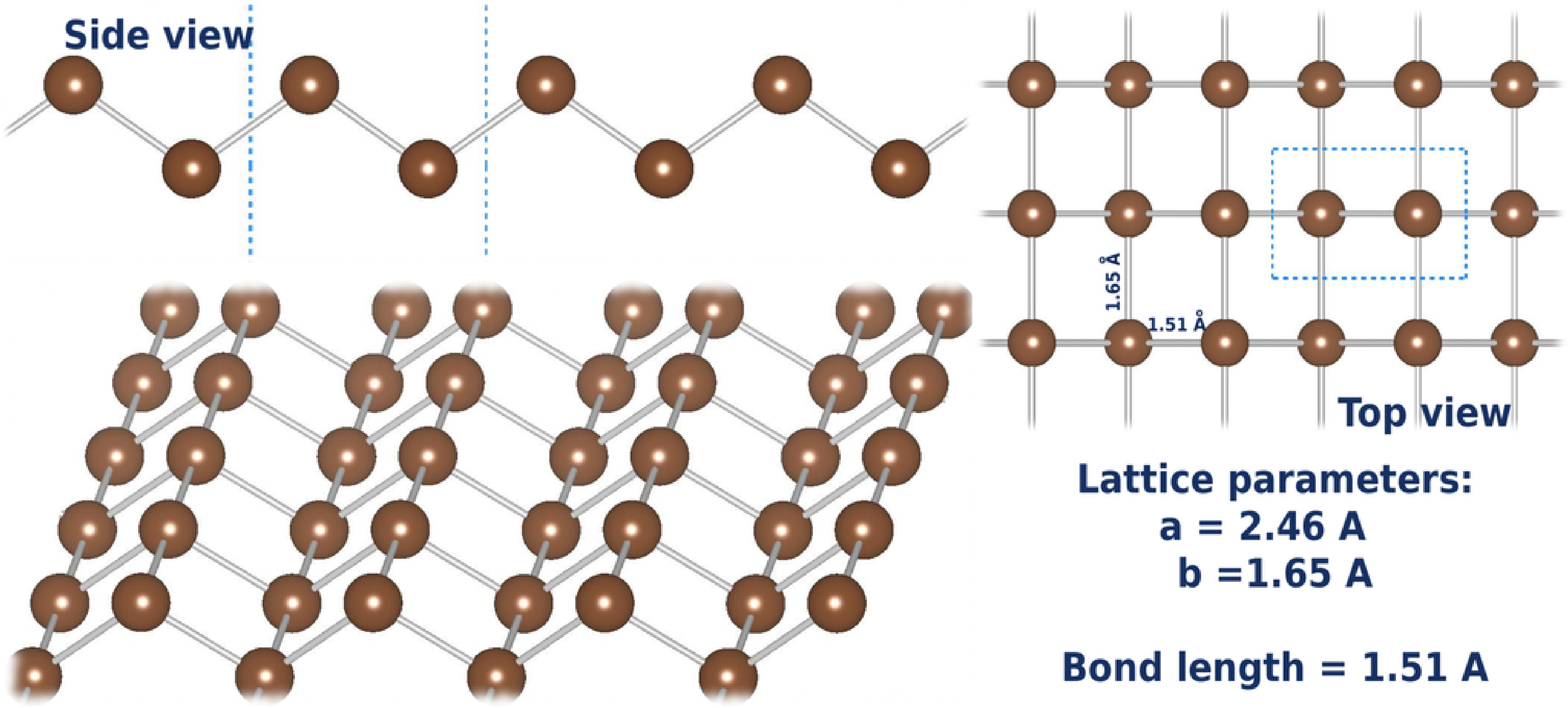}
		\line(0,1){140}
		\includegraphics[width=.40\textwidth]{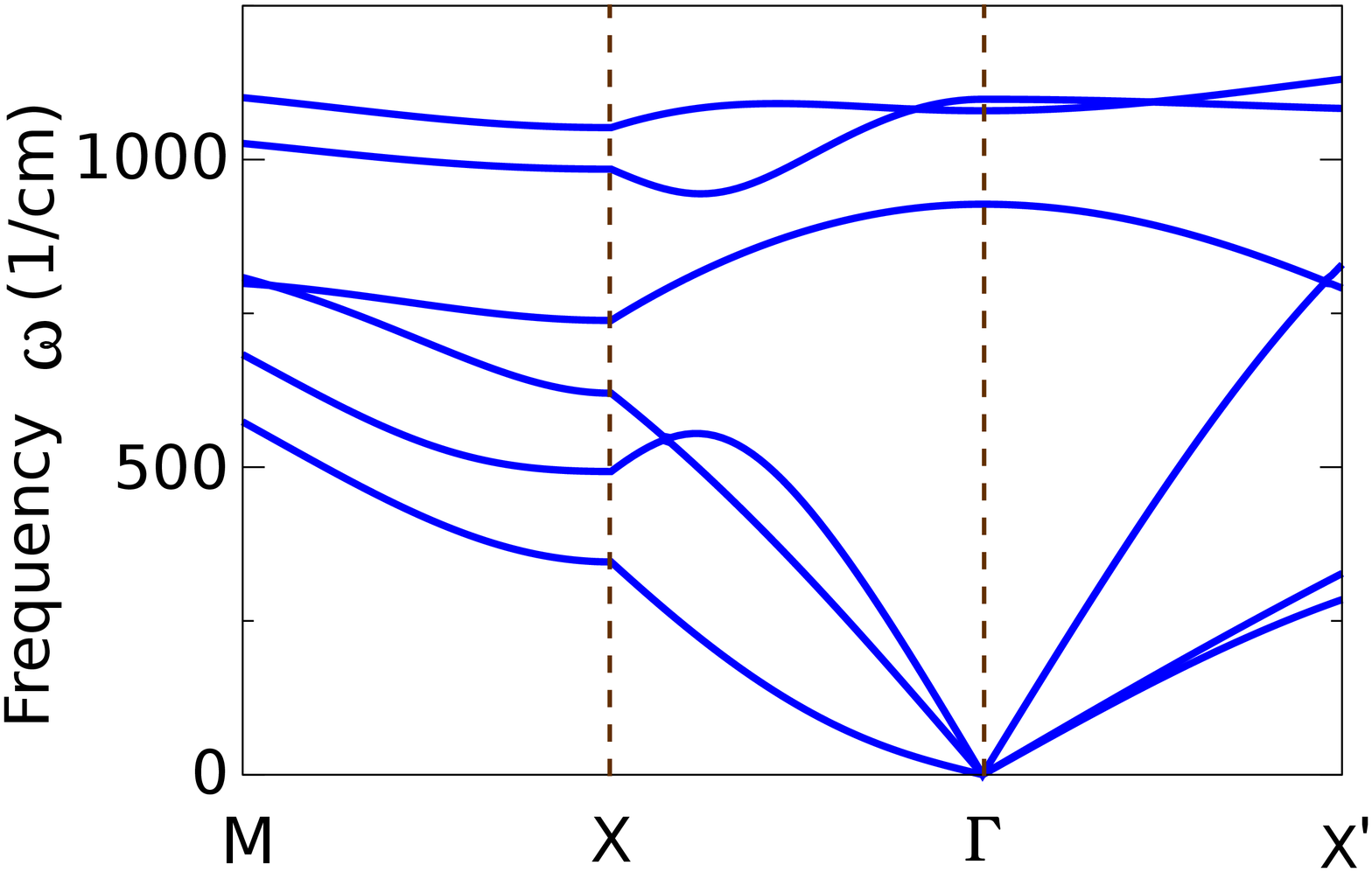}\\
		~ ~ ~ (a) \hskip 9cm (b)
	\end{center}
	\caption{(a) Optimized structure of ZzC. The name zigzag is evident from
	the side view. The dotted blue line indicate the primitive unit cell
	which has the parameters $a= 2.46$ {\AA}, $b=1.65$ {\AA} and $\gamma =
	90^\circ$. The primitive cell has two carbon atoms with the bond length
	of 1.51 {\AA}. (b) The phonon dispersion relation is shown along the
	symmetry points of Brillouin zone. The absence of soft modes confirms
	the stability.}
	\label{fig:struct}
\end{figure*}

Figure \ref{fig:struct}(a) shows the optimized crystal structure along with its
lattice parameters and bond lengths. For sake of clarity various views are
depicted, which also make the name zigzag carbon evident. The unit cell belongs
to $C_{2h}$ point group of monoclinic crystal, with lattice parameters of
$a=2.46$ {\AA}, $b=1.65$ {\AA}. As seen from the top view the bond length along
the zigzag direction (1.51 {\AA}) is smaller than that along non-zigzag
direction (1.65 {\AA}). These numbers are consistent with Wen {\it et al}
\cite{CHEM:CHEM200903128} (1.50 and 1.63 {\AA} respectively; bearing in mind the
choice of different treatment on exchange correlation functional).  At this
point we recall that the C-C bond length is 1.42 {\AA} and 1.62 {\AA} in graphene and SqC (planar square carbon allotrope \cite{pujari:sqc}) respectively. It is thus intuitive to think that binding energy of ZzC may lie
in between the two. Indeed, ZzC has the binding energy of 7.46 eV per atom which
is higher than SqC (6.7 eV per atom) but not as high as graphene (9.6 eV per
atom). However binding energy alone  cannot serve as the stability criterion.
The stability of ZzC becomes evident conclusively by analysing the phonon
dispersion spectrum (Figure \ref{fig:struct}(b)). Obviously, the lack of any
imaginary phonons in entire Brillouin zone strongly underlines the stability of
ZzC. The phonon spectrum has total of six branches, three acoustic and three
optical (out-of-plane, in-plane transverse and in-plane longitudinal for both
types). It is interesting to note that although there is no band gap between
acoustic and optical branches, an indirect band gap of 18.9 cm$^{-1}$ exists
between the out-of-plane optical mode and in-plane transverse optical mode.
Thus, from phonon dispersion it becomes evident that ZzC is indeed a stable
material with a high binding energy.
\subsection{Electronic structure}
\begin{figure*} 
	\begin{center}
		\includegraphics[width=.52\textwidth]{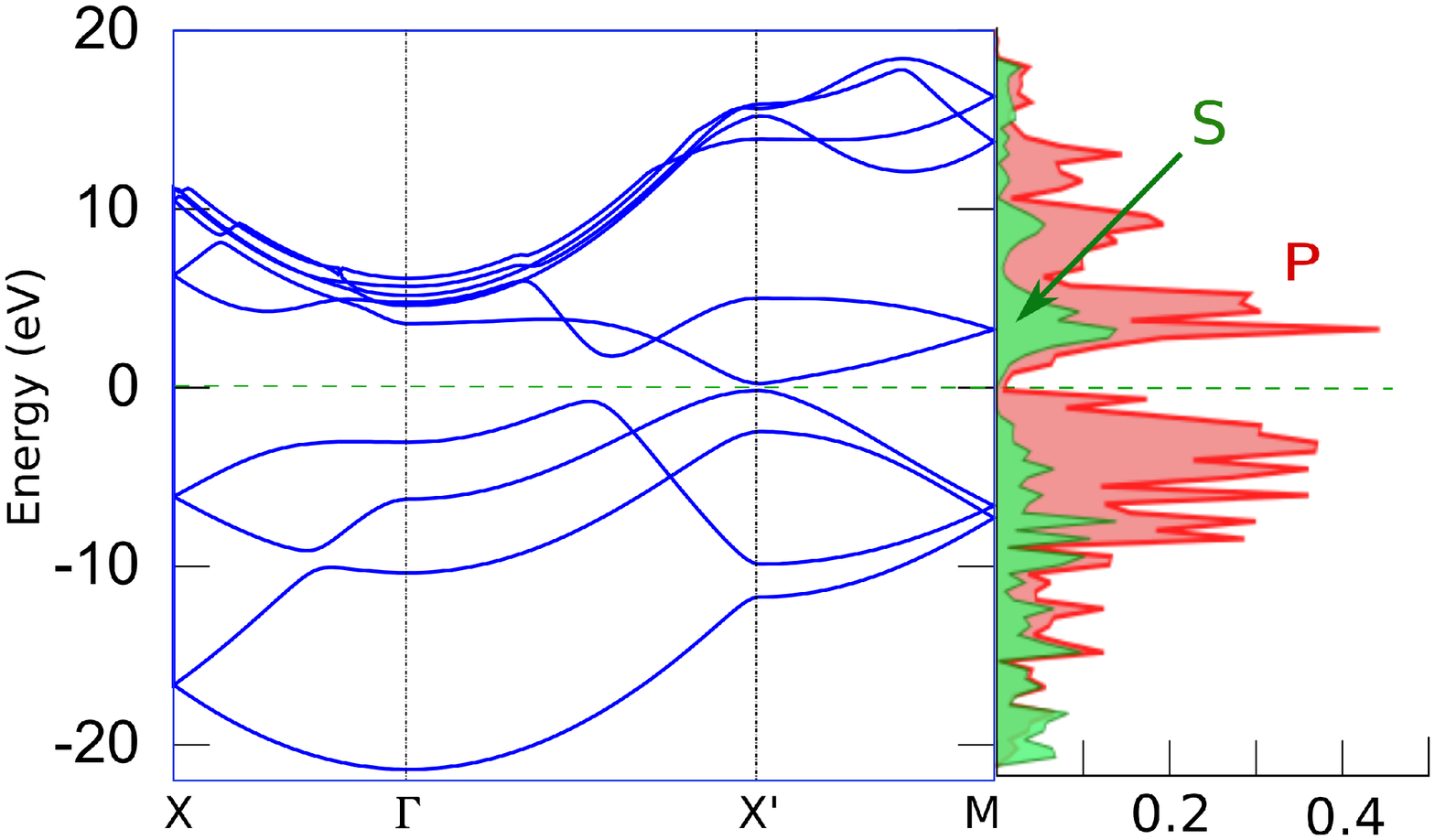}
		\includegraphics[width=.2\textwidth]{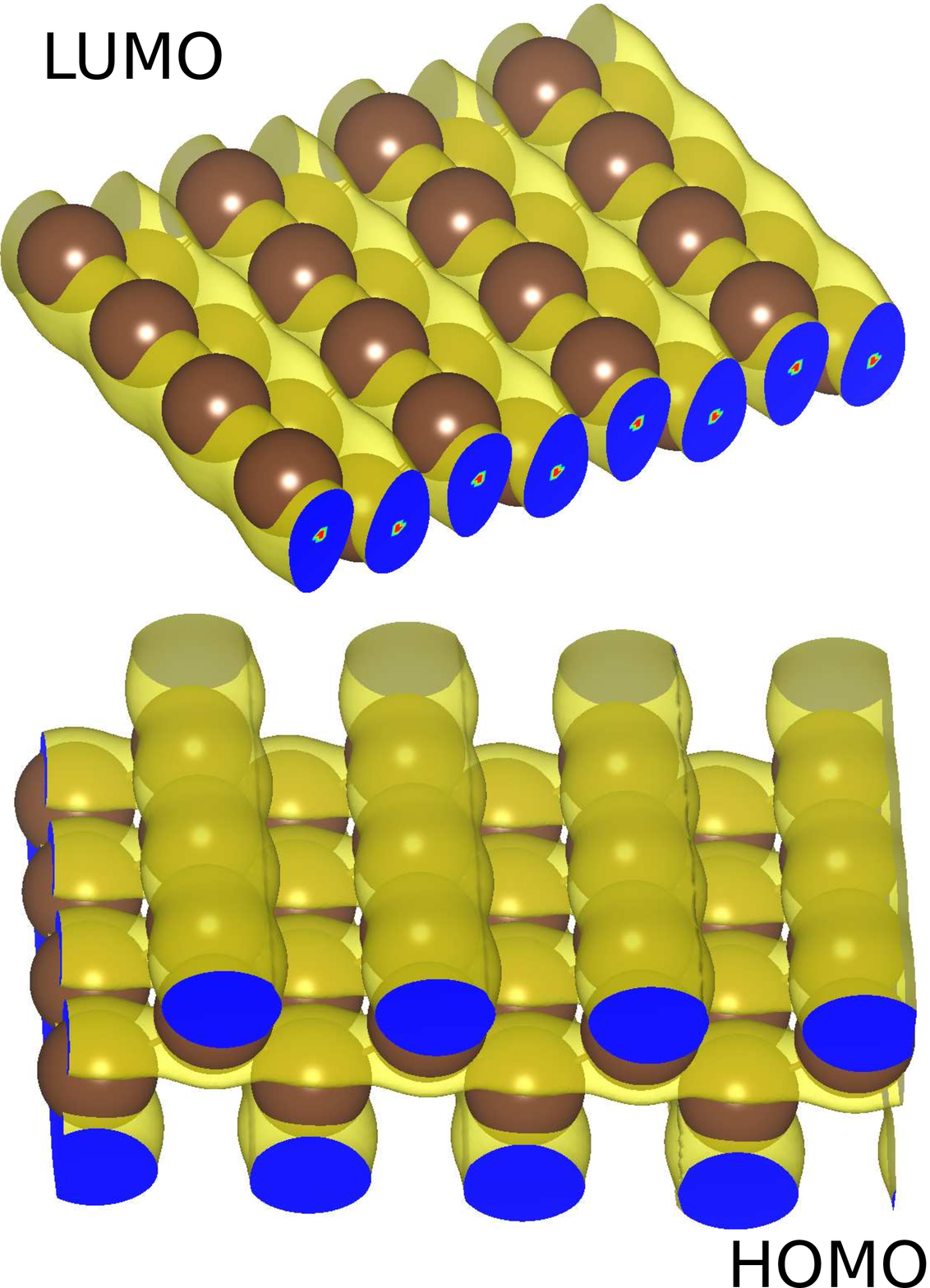} \\
		(a) \hskip 6cm (b)
	\end{center}
	\caption{(a)Band structure and the DOS of ZzC. DOS are resolved as per
	their angular momentum components ($s$ and $p$) and are plotted using
	the electronic smearing of 0.1 eV for better visualization. A small band
	gap of 0.39 eV is seen at the $X'$ point. (b) Charge density of highest
	occupied and lowest unoccupied bands show the declocalization
	perpendicular to zigzag direction. }
	\label{fig:BS} 
\end{figure*}

After validating the stability we now discuss the electronic structure of ZzC.
As seen from the band structure (Figure \ref{fig:BS}(a)) ZzC shows a small band
gap of 0.39 eV at the $X'$ point of Brillouin zone. The band gap at $\Gamma$
point is rather high (7.8 eV). We also note that ZzC does not show any magnetic
character. Figure \ref{fig:BS}(b) also shows the charge densities of highest
occupied and lowest unoccupied bands. Highest occupied band which is primarily
$p$-type shows the stacking of baguette-shaped charge density, while unoccupied
band completely embed the atoms within individual planes. It is interesting to
note that both states show the delocalization perpendicular to the
zigzag direction. Looking at the delocalized nature of bands and bearing in mind
that DFT tends to underestimate the band gap, one may speculate that the ZzC may
act as a semiconductor with peculiar conductance perpendicular  to zigzag
direction.

\subsection{Reaction pathway}

\begin{figure*}
	\begin{center}
		\includegraphics[width=.8\textwidth]{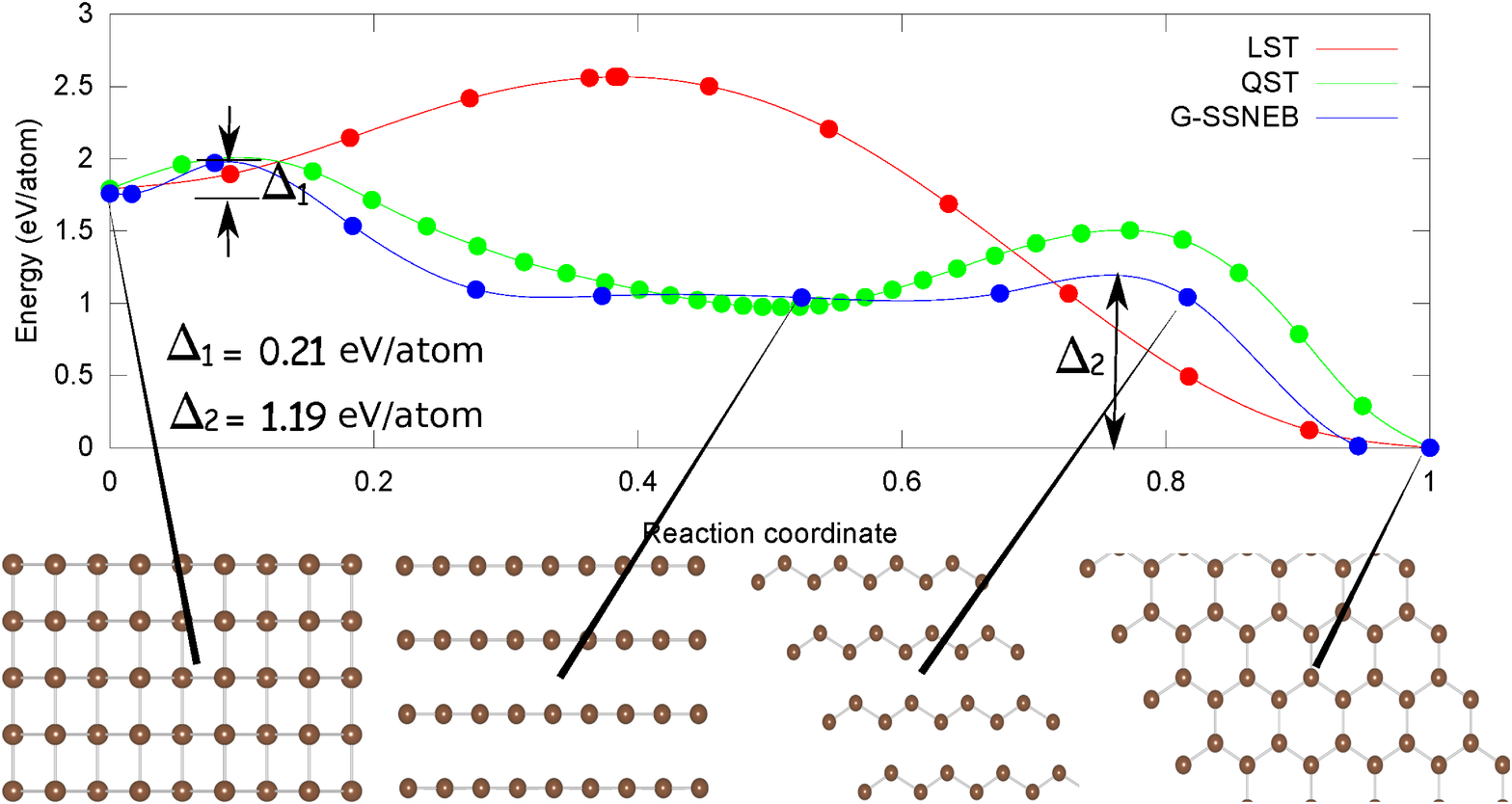}
	\end{center}
	\caption{Reaction pathways from ZzC to graphene using LST, QST and
	G-SSNEB methods (in the last one we used two climbing images because of
	two reaction barriers).}
	\label{fig:lstqst}
\end{figure*}

LST, QST and G-SSNEB pathways along with initial, transition, intermediate and
final states, are presented in Figure 3. LST method does not have the
intermediate state found by QST and G-SSNEB independetly. It is explained by
non-iterative nature of LST: it represents only the shortest pathway between
reactant and product, and is useful only as a quick and rough approximation.
Presence of the intermediate state,  carbyne (weakly binded carbon chains), means a
multistep reaction: along the pathway to the more stable graphene allotrope ZzC
dissociates to carbyne. Carbyne structure (Figure 3) is composed of linear
carbon chains\cite{Kim2009} weakly binded by dispersion
interaction. Our results are in good agreement with parameters of betta-carbyne (or polycumulune type, cumulated double bonds) in Heinmann's work \cite{Heimann1983}. In particular double C=C bond is equal 1.28 {\AA} in both cases, which is in agreement with the Steichen's pricinple \cite{Stoicheff1962}. Interchain distances are 3.7 and 2.9 {\AA} in our and Heinmann's work accordingly. The bigger value in our case is due to different unit cells in the works. In our caclulation we used artificially small unit cell for carbyne: there are only two carbon atoms, whereas in Heinmann's work n-atoms cells ($n>>2$) are presented. Bigger cell allows more energetically favourable packing, which leads to higher interaction, thus smaller interchain distance. That difference also explains different transition temperatures from graphene to carbyne: in our calculation reaction barrier from graphene to carbyne is 1.19 eV/atom, that corresponds to about 4600 K (the method of temperature calculation is described in the next paragraph), whereas in the work of Nelson etc. \cite{Nelson1972} graphene was heated above 2600 K. The planar structure of carbyne is most likely an artefact
caused by construction of primitive cell (to consider possibility of non-planar
structure supercell in the orthogonal direction to the chains should be
constructed). In reality final structure could be a layer or even clusters of
amorphous carbon composed of crosslinked linear chains \cite{Kim2009,
Lebedev2000}. But this part of modeling lies beyond the scope of this article.

To answer the question of stability of ZzC structure we calculated transition temperature corresponding to reaction barrier using rough approximation of atoms as independent simple harmonic oscillators having energy $3kT$ each. Thus, a rough value of the transition temperature is $E/(3k)$, where $E$ and $k$ are energy of reaction barrier per atom and Boltzmann constant accordingly. For transition from ZzC to carbyne it corresponds to around 815 K. Thus ZzC structure is a pretty stable composition.\\

\subsection{Hydrogenation}
At this stage we wish to point out an interesting observation. Unlike graphene
or SqC \cite{pujari:sqc}, ZzC responds to hydrogenation in qualitatively
different manner. We recall that upon hydrogeation graphene forms {\it
graphane}, while SqC  accommodates hydrogens by forming hexacoordinated bonds.
In contrast, we observed that upon hydrogenation, ZzC breaks down and forms
stacks of polyacetylene (-[CH]$_n$-). The individual chains of polymers were
spatially separated by more than 3.5 {\AA}, ruling out the van der Waals
interactions. We verified the absence of any such interactions by enabling a
correction term to the exchange correlation functional, to take into account the
long range the dispersion.\cite{london1,london2} Yet, such different response to
hydrogenation is understandable by analyzing the nature of electronic structure.
While graphene has one double bond, SqC has electron deficient bonding. On the
other hand, all bonds in ZzC are single bonds and are completely saturated.  By
disintegrating into polyacetylene carbon atoms gain significant binding energy
(from 7.46 eV to 8.31 eV per atom). Thus although the ZzC is extremely stable,
we speculate that it can be easily converted into stacks of polyacetylene by
hydrogenation.

\section{Conclusions}
We performed the full unit cell optimization on
zigzag carbon and found out that the ZzC may act as the semiconductor with
valance and conductance bands peculiarly delocalized only perpendicular to the
zigzag direction. Our calculations revealed that with binding energy of 7.46 eV
per atom ZzC is stable, which we  also verified  using lattice
dynamical calculations. We also observed that ZzC breaks into chains of
polyacetylene upon hydrogenation. Finally our reaction path from ZzC to graphene using QST and G-SSNEB
showed that ZzC transforms into carbyne with reaction barrier 0.21 eV/atom, roughly corresponding 815 K transtion temperature. We believe many other exotic 2D carbon allotropes have the same fate: breaking apart to carbon chains with following amorphous phase formation.

\bibliographystyle{unsrt} \bibliography{biblio}

\end{document}